\documentclass[traditabstract,longauth]{aa} % for the long lists of affiliations 

\usepackage{graphicx}
\usepackage{txfonts}
\begin{document}
\title{{\it Herschel} ATLAS:\thanks{{\it Herschel} is an ESA space observatory
    with science instruments provided by European-led Principal
    Investigator consortia and with important participation from
    NASA.}  The angular correlation function of submillimetre galaxies at
  high and low redshift.  }

\author{  
S.J. Maddox\inst{1} 
\and  
L. Dunne\inst{1} 
\and
E. Rigby\inst{1}
\and  
S. Eales\inst{2}
\and
A. Cooray\inst{9}
\and
D. Scott\inst{18} 
\and 
J.A. Peacock\inst{19} 
\and 
M. Negrello\inst{16}
\and
D.J.B. Smith\inst{1}
\and
D. Benford\inst{15} 
\and
A. Amblard\inst{9}
\and
R. Auld\inst{2}
\and
M. Baes\inst{3}
\and 
D. Bonfield\inst{4}
\and
D. Burgarella\inst{5}
\and
S. Buttiglione\inst{6}
\and
A. Cava\inst{7}$^,$\inst{20}
\and
D. Clements\inst{8}
\and
A. Dariush\inst{2}
\and
G. de Zotti\inst{6}
\and
S. Dye\inst{2}
\and
D. Frayer\inst{10}
\and
J. Fritz\inst{3}
\and
J. Gonzalez-Nuevo\inst{11}
\and
D. Herranz\inst{12}
\and
E. Ibar\inst{13}
\and
R. Ivison\inst{13}
\and
M.J. Jarvis\inst{4}
\and
G. Lagache\inst{14}
\and
L. Leeuw\inst{15}
\and
M. Lopez-Caniego\inst{12}
\and
E. Pascale\inst{2}
\and
M. Pohlen\inst{2}
\and
G. Rodighiero\inst{6}
\and
S. Samui\inst{11}
\and
S. Serjeant\inst{16}
\and
P. Temi\inst{15}
\and
M. Thompson\inst{4}
\and
A. Verma\inst{17}
}

\institute{
School of Physics and Astronomy, University of Nottingham,
University Park, Nottingham NG7 2RD, UK
\and
School of Physics and Astronomy, Cardiff University,
  The Parade, Cardiff, CF24 3AA, UK
\and
Sterrenkundig Observatorium, Universiteit Gent, Krijgslaan 281 S9,
B-9000 Gent, Belgium
\and
Centre for Astrophysics Research, Science and Technology Research
Centre, University of Hertfordshire, Herts AL10 9AB, UK
\and
Laboratoire d'Astrophysique de Marseille, UMR6110 CNRS, 38 rue F.
Joliot-Curie, F-13388 Marseille France
\and
University of Padova, Department of Astronomy, Vicolo Osservatorio
3, I-35122 Padova, Italy
\and
Instituto de Astrof\'{i}sica de Canarias, C/V\'{i}a L\'{a}ctea s/n, E-38200 La
Laguna, Spain
\and
Astrophysics Group, Imperial College, Blackett Laboratory, Prince
Consort Road, London SW7 2AZ, UK
\and
Dept. of Physics \& Astronomy, University of California, Irvine, CA 92697, USA
\and
Infrared Processing and Analysis Center, California Institute of
Technology, 770 South Wilson Av, Pasadena, CA 91125, USA
\and
Scuola Internazionale Superiore di Studi Avanzati, via Beirut 2-4,
34151 Triest, Italy
\and
Instituto de F\'isica de Cantabria (CSIC-UC), Santander, 39005, Spain
\and
UK Astronomy Technology Center, Royal Observatory Edinburgh, Edinburgh, EH9 3HJ, UK
\and
Institut d'Astrophysique Spatiale (IAS), Bâtiment 121, F-91405 Orsay, France; and Université Paris-Sud 11 and CNRS (UMR 8617), France
\and
Astrophysics Branch, NASA Ames Research Center, Mail Stop 245-6, Moffett Field, CA 94035, USA
\and
Dept. of Physics and Astronomy, The Open University, Milton Keynes, MK7 6AA, UK
\and
Oxford Astrophysics, Denys Wilkinson Building, University of Oxford,
Keble Road, Oxford, OX1 3RH
\and
University of British Colombia, 6224 Agricultural Road, Vancouver, BC
V6T 1Z1, Canada
\and
SUPA, Institute for Astronomy, University of Edinburgh, Royal Observatory Edinburgh, Edinburgh, EH9 3HJ, UK
\and
Departamento de Astrof{\'\i}sica, Universidad de La Laguna (ULL),
E-38205 La Laguna, Tenerife, Spain
}

%   \institute{
%     School of Physics and Astronomy, University of Nottingham
%     \email{steve.maddox@nottingham.ac.uk}
%     \and
%     Dept of Physics and Astronomy, University of Cardiff
%     \and
%     other places
%   }

   \date{} %Received September 15, 1996; accepted March 16, 1997}

% \abstract{}{}{}{}{} 
   % 5 {} token are mandatory
 
  \abstract
  % context heading (optional)
  % {} leave it empty if necessary {} 
{

We present measurements of the angular correlation function of
galaxies selected from the first field of the H-ATLAS survey. Careful
removal of the background from galactic cirrus is essential, and
currently dominates the uncertainty in our measurements. For our
$250\ \mu$m-selected sample we detect no significant clustering,
consistent with the expectation that the $250\ \mu$m-selected sources
are mostly normal galaxies at $z\lesssim 1$. For our $350\ \mu$m and
$500\ \mu$m-selected samples we detect relatively strong clustering
with correlation amplitudes $A$ of 0.2 and 1.2 at $1'$, but with
relatively large uncertainties. For samples which preferentially
select high redshift galaxies at $z\sim2-3$ we detect significant
strong clustering, leading to an estimate of $r_0 \sim 7-11 \,h^{-1}$
Mpc. The slope of our clustering measurements is very steep,
$\delta\sim2$. The measurements are consistent with the idea that
sub-mm sources consist of a low redshift population of normal galaxies
and a high redshift population of highly clustered star-bursting
galaxies.

}
%AA/2010/14663

  % aims heading (mandatory) {}
  % methods heading (mandatory) {}
  % results heading (mandatory) {} 
  % conclusions heading (optional), leave it empty if necessary  {}


   \maketitle
%
%_____e___________________________________________________________

\section{Introduction}

Submillimetre (sub-mm) selected galaxy samples provide a unique way to
trace obscured star formation out to high redshifts (Blain et
al. \cite{blain02}). Models for the evolution of such populations
disagree on the nature of the sub-mm sources at high redshifts, with
some claiming that they are massive galaxies in the process of forming
most of their stellar mass (Granato et al. 2004, Narayanan et
al. 2009, Dav\'e et al. 2010) while others model them as lower mass
sources undergoing bursts of star formation with a top heavy IMF
(Baugh et al. 2005, Lacey et al. 2009). While evidence on individual
sources largely supports a massive halo scenario (Dunne, Eales \&
Edmunds 2003, Swinbank et al. 2008, Michalowski, Hjorth \& Watson
2010), the best way to measure the statistical halo properties of this
population is to measure their clustering. The three-dimensional
clustering of sub-mm galaxies provides information about the
dark-matter halos that they populate, but direct measurements need
distance estimates for each galaxy, which we do not have for our
sample. The angular clustering can be measured for flux-limited
samples but, in order to compare with models, predictions are required
for both the $n(z)$ of flux limited samples and the intrinsic 3-d
clustering of the galaxies. The sub-mm colours depend on the source
redshift, and so selecting on colour can preferentially select high or
low redshift samples (see e.g. Amblard et al. 2010).

Previous attempts to measure the clustering of sub-mm sources have
been mostly based on catalogues which cover only very small areas.
Despite predictions that sub-mm galaxies should have high spatial
clustering, their $n(z)$ is broad and so previous work has had limited
success in detecting a significant angular clustering signal (Blain et
al. 2004, Scott et al. 2006, Wei\ss\ et al. 2010). A more recent
approach has been to measure the power spectrum of larger area sub-mm
maps from BLAST (Viero et al. \cite{viero}, Devlin et
al. \cite{blast}). This analysis has found significant evidence for
clustering, although at relatively low amplitude.

The {\it Herschel} ATLAS (H-ATLAS) (Eales et al. \cite{eales}) will survey
over 550 deg$^2$ in 5 wavebands at 100, 160, 250, 350 and $500
\ \mu$m. One field covering $\sim 4^\circ\times4^\circ $ degrees was
observed during the science demonstration phase of the mission, and
has produced a catalogue of $\sim 6600$ sources with significance $>5
\sigma$ above the combined instrumental and confusion noise. This
represents roughly 1/30 of the final H-ATLAS data-set.  In this paper
we present measurements of the angular correlation function of five
flux and colour-selected samples of the H-ATLAS sources.

% The paper is organized as follows: section 2 presents the data, with
%brief descriptions of the source map making and source detection;
% section 3 describes how we measure the angular correlation function;
% section 4 compares the measurements to other data and model
% predictions and discusses our conclusions.

%__________________________________________________________________

\section{H-ATLAS data and source catalogues}
\label{sec:2}

\begin{figure}
\hspace{10mm}\includegraphics[width=0.39\textwidth]{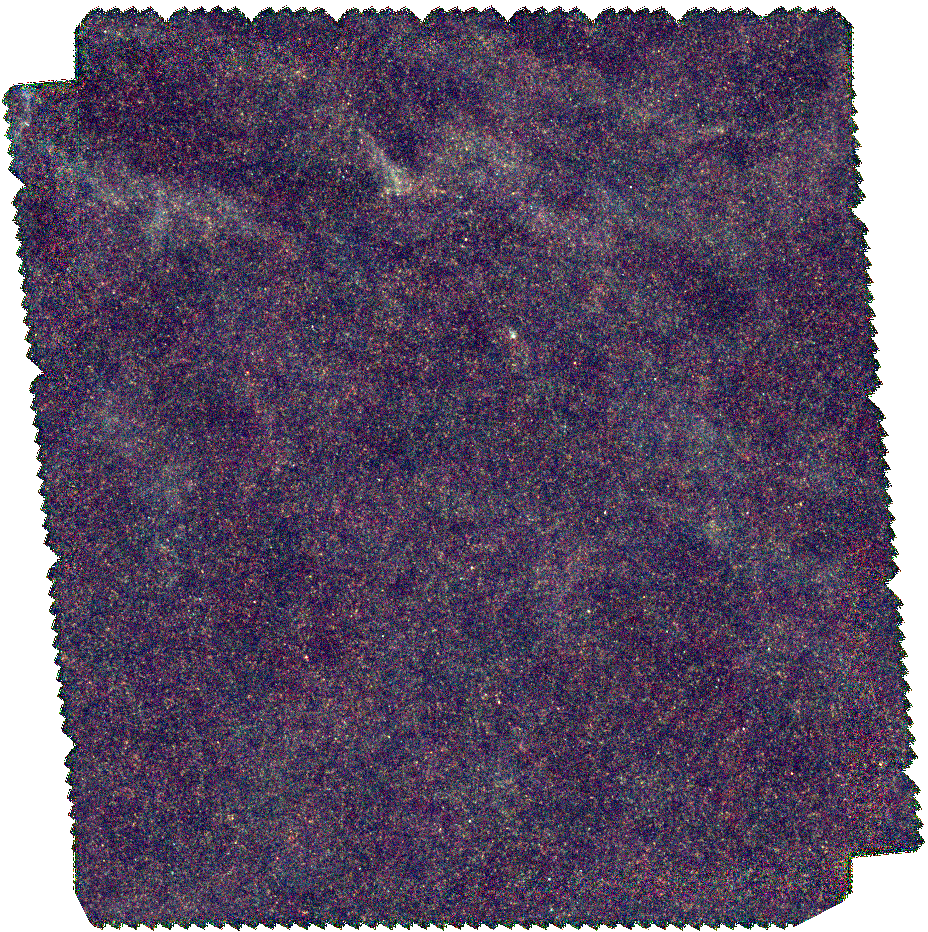}
%\hspace{10mm}\includegraphics[width=0.4\textwidth]{big_colour_image.png}
% figure caption is below the figure
\caption{False-colour image combining the SPIRE 250, 350 and
  500$\ \mu$m maps as blue, green and red respectively. The ragged
  edges show the individual scan-legs of the two scans.  Galactic
  cirrus can be seen as patchy blue wisps over the field. }
\label{fig:1}       % Give a unique label
\end{figure}

H-ATLAS uses parallel scan mode observations performed with the ESA
{\it Herschel} Space Observatory (Pilbratt et al. \cite{herschel}),
providing data simultaneously from both the PACS (Poglitsch et
al. \cite{pacs}) and SPIRE (Griffin et al. \cite{spire}) instruments.
The time-line data are reduced using {\tt HIPE}. Maps are produced
from the SPIRE data using a naive mapping technique after removing
instrumental temperature variations from the time-line data (Pascale
et al. in prep). Noise maps are generated by using the two
cross-scan measurements to estimate the noise per detector pass, and
then for each pixel the noise is scaled by $N_{passes}^{1/2}$, where
$N_{passes}$ is the number of detector passes.  A false-colour image
combining the SPIRE 250, 350 and 500$\ \mu$m maps is shown in
Fig.~\ref{fig:1}. The PACS H-ATLAS maps currently yield only a
few hundred sources, which is insufficient for us to attempt a
clustering measurement. The clustering in the PACS bands will be
investigated in a future paper using more data.

\begin{figure}
%\hspace{10mm}
\includegraphics[width=0.5\textwidth,angle=-90]{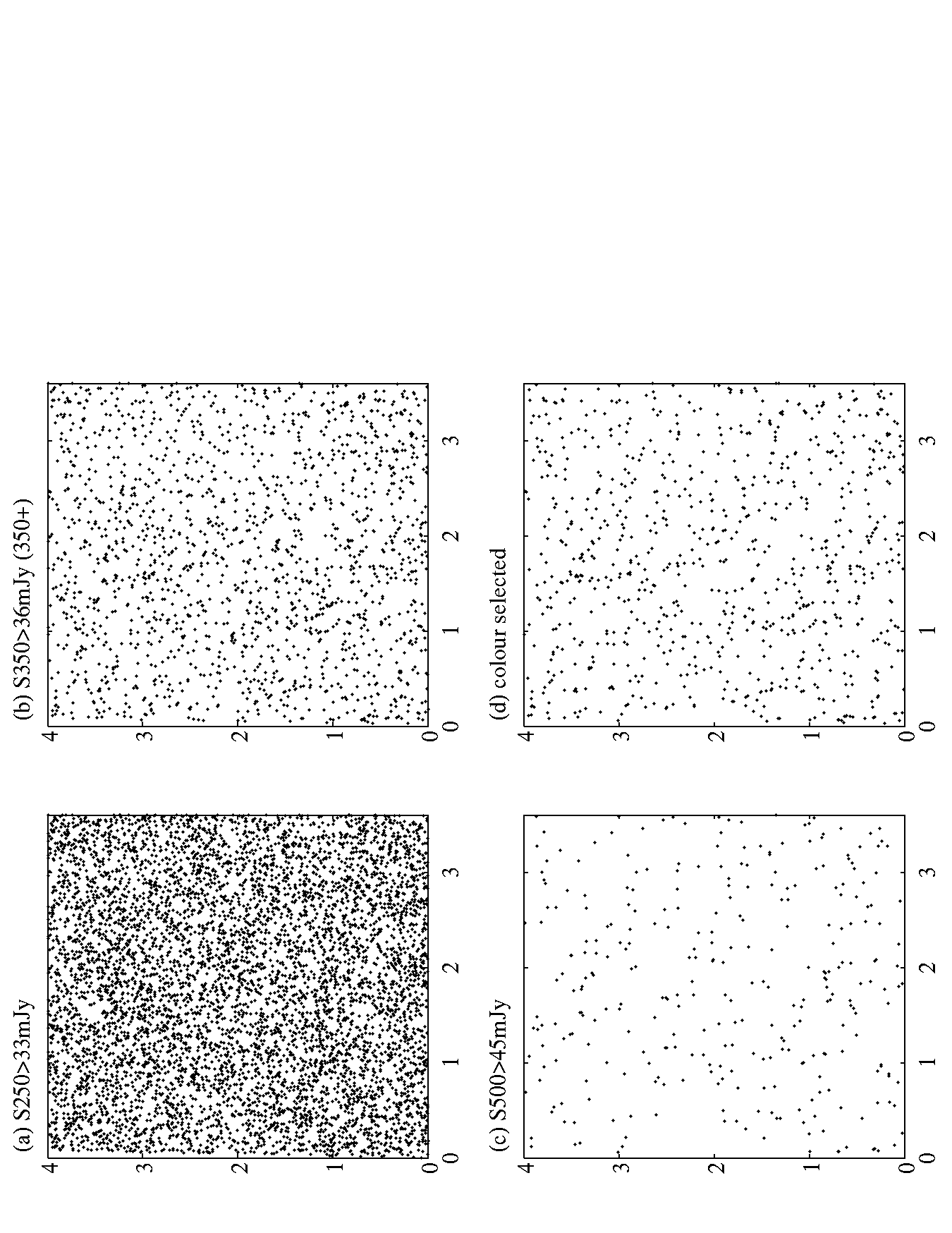}
\caption{The positions of sources in four subsamples (a)
  $S_{250}>33$mJy; (b) $S_{350}>36$mJy and 3$\sigma$ in $S_{250}$ and
  $S_{500}$ (the 350+ sample); (c) $S_{500}>45$mJy; (d) the
  colour selected sample $S_{500}/S_{250}>0.75$.  }
\label{fig:2}       % Give a unique label
\end{figure}

Sources were identified in the SPIRE maps using a Multi-band Algorithm
for source eXtraction (MADX, Maddox et al. in prep).  First a local
background is estimated from the peak of the histogram of pixel values
in $30\times 30$ blocks of pixels. This corresponds to $2.5'$ for the
250$\ \mu$m map, and $5'$ for the 350 and 500$\ \mu$m maps. The
background at each pixel is then estimated using a bi-cubic
interpolation between the coarse grid of backgrounds, and this value
subtracted from the pixel.  The filter scale was chosen to be as large
as possible while still following the variations in the cirrus
background. Since the local background is estimated from the peak of
the flux histogram, it is insensitive to the presence of resolved
sources within the background block, so long as they do not cover a
significant number of pixels in the block. This approach should remove
the local background without removing flux from the resolved sources,
and so should be less susceptible to removing real structure in the
source distribution compared to standard Fourier filtering approaches.

The background subtracted maps are then filtered by the estimated PSF,
including a local inverse variance weighting.  The maps from all three
bands are then combined with weights set by the local inverse
variance, and also the prior expectation of the SED of the
galaxies. We tried a flat-spectrum prior, where equal weight is given
to each band and also 250$\ \mu$m weighting, where only the
250$\ \mu$m band was included. At the depth of the filtered maps
source confusion becomes an issue in the longer wavelength bands, and
the higher resolution of the 250$\ \mu$m maps outweighs the
signal-to-noise gain from adding in the other bands. The current
catalogues use the 250$\ \mu$m-only prior and we will revisit this
issue in future data releases.

All local peaks are identified in the combined PSF filtered map as
potential sources, and a Gaussian is fitted to each peak to give
estimates of the position at the sub-pixel level and the point source
flux.  The flux densities in other bands are estimated by using a
bi-cubic interpolation to the position given by the combined map.  To
produce a catalogue of reliable sources, we select only sources that
are detected at the 5-$\sigma$ level in any of the bands.  In
calculating the $\sigma$ for each source, we use the relevant noise
map, and add the confusion noise to this in quadrature. The average
1-$\sigma$ instrumental noise values in the PSF-filtered maps are 4, 4
and 5.7mJy beam$^{-1}$ respectively in the 250, 350 and 500$\ \mu$m
bands. We estimated the confusion noise from the difference between
the variance of the maps and the expected variance due to instrumental
noise, and find that the 1-$\sigma$ confusion noise is 5, 6 and 7 mJy
beam$^{-1}$ at 250, 350 and 500$\ \mu$m. The resulting total
5-$\sigma$ limits are 33, 36 and 45mJy beam$^{-1}$ (Rigby et
al. in prep). Source counts from these catalogues are analysed by
  Clements et al. (\cite{counts}), and are found to be consistent with
  previous measurements in these wave-bands (Patanchon et
  al. \cite{blastc}).

We have selected five samples to use for our current clustering
analysis. The first three use simple flux density cuts, as given in
Table~\ref{tab:1}. The fourth and fifth samples are as defined by
Amblard et al. (\cite{Amblard}), who use the H-ATLAS colours to
estimate redshift distributions. The fourth sample, which we call
$350+$ is $>5\sigma$ at 350$\mu$m with an extra constraint that the
sources must also be detected at more than $3\sigma$ in the 250$
\ \mu$m and 500$ \ \mu$m bands, and the fifth sample adds a further
constraint that the ratio $S_{500}/S_{250}>0.75$. Requiring a
detection at 500$ \ \mu$m tends to select higher redshift galaxies
compared to a simple 350$ \ \mu$m selection, and Amblard et
al. estimate that the mean redshift of the $350+$ sample is
$2.2\pm0.6$. The $S_{500}/S_{250}>0.75$ colour selection pushes to an
even higher redshift, of $2.6\pm0.3$.  The positions of sources in
four of the sub-samples are shown in Fig.~\ref{fig:2}. 

\begin{table}
      \caption[]{Subsamples used to measure $w(\theta)$, and best-fit
        power-law parameters. $N$ is the number of sources in each
        sample. $A$ is the amplitude at $1'$ and $\delta$ is the
        power-law slope. $A_{0.8}$ and  $A_{2.0}$ are the amplitudes
        at $1'$ with the slopes fixed at $0.8$ and $2.0$
        respectively. 
      \label{tab:1}
         $\begin{array}{lccccc}
            \hline
            \noalign{\smallskip}
            {\rm Sample } &  N & A &
            \delta & A_{0.8} & A_{2.0} \\
            \noalign{\smallskip}
            \hline
            \noalign{\smallskip}
            S_{250}>33 & 6317 & -0.01\pm 0.07& 1.7\pm 0.2 &-0.00 & -0.01\\
            S_{350}>36 & 2754 & 0.20\pm 0.07 & 2.0\pm 0.2 & 0.11 & 0.20\\
            S_{350}>36^{\ast} & 1633 & 0.50 \pm 0.09 & 2.8\pm 0.5 & 0.21 & 0.50\\
            S_{500}>45 & 304  & 1.24\pm 1.6  & 2.4\pm 1.3 & 0.51 & 1.24\\
            S_{500}/S_{250}>0.75 & 808 & 0.92 \pm 0.3 & 2.1\pm 0.5 & 0.38 & 0.92\\
            \noalign{\smallskip}
            \hline
         \end{array} $\\
         {\small $^{\ast}$ this is the $350+$ sample which has the
           additional constraint that source must be detected at $>3
           \sigma$ in the other two bands. }}
   \end{table}
%__________________________________________________________________

\section{Measuring $w(\theta)$ }
\label{sec:3}

We measure the correlation function by counting pairs in the data as a
function of angular separation and comparing to the number of pairs in a
random catalogue with similar boundaries and selection effects.  The pair
counts are combined to estimate the correlation function, $w(\theta)$, using
the Landy \& Szalay estimator (\cite{LS})
\begin{equation}
w(\theta) = \frac{DD-2DR+RR }{RR} .
\end{equation}
Here $DD$ is the number of data-data pairs, $DR$ is the number of
data-random pairs and $RR$ is the number of random-random pairs, each
at separation $\theta$.  

The random catalogues were generated to follow the sensitivity limit
of the actual data selection. This means that any non-uniformities in
the data due to variation in signal-to-noise should not be imprinted
on the clustering signal.  To relate the noise at a pixel to the
expected number density of sources we generated random fluxes which
match the observed count slope (Clements et al. \cite{counts}),
perturbed them by a Gaussian deviate with standard deviation equal to
the local noise estimate, and then kept the random source if it was
brighter than the chosen flux limit. In practice the noise maps are
uniform enough that using uniform random catalogues makes no
significant difference to the results.

The clustering measurements are sensitive to the correct removal of
the spatially varying cirrus background, as well as the unresolved
background of faint sources, which are also likely to be strongly
clustered. We have investigated the stability of the measurements by
masking areas around the brightest patches of cirrus. This had little
effect on the measurements indicating that our cirrus removal is
effective.  Increasing the scale of background filtering to 60 and 120
pixels produces a much larger clustering signal. A visual inspection
of the source positions makes it clear that the excess structure in
the source distribution is correlated with the pattern of cirrus
emission, and so is likely to be spurious signal caused by
insufficient background subtraction. 

A potential concern when using such a small scale to remove the
background is that some real clustering may have been removed. This
was tested by using clustered source positions to create simulated
maps, which include cirrus background estimated from the IRAS maps
(Schlegel Finkbeiner \& Davis \cite{sfd}) and the same noise and
coverage maps as the real data. The background was then filtered and
sources extracted using the MADX algorithm as for the real data, and
the clustering of the resulting sources measured. The clustering
amplitude recovered from the simulations varied with background
subtraction scale in a similar way to the real data. The correct
amplitude was recovered using 30 pixels; using 15 pixels
underestimated the amplitude by $\sim 10$\%.  We therefore believe
that our background subtraction has removed the effect of cirrus on
the source clustering, yet has not removed true structure in the
source distribution.

\begin{figure*}
 \centerline{\includegraphics[width=0.8\textwidth,angle=0]{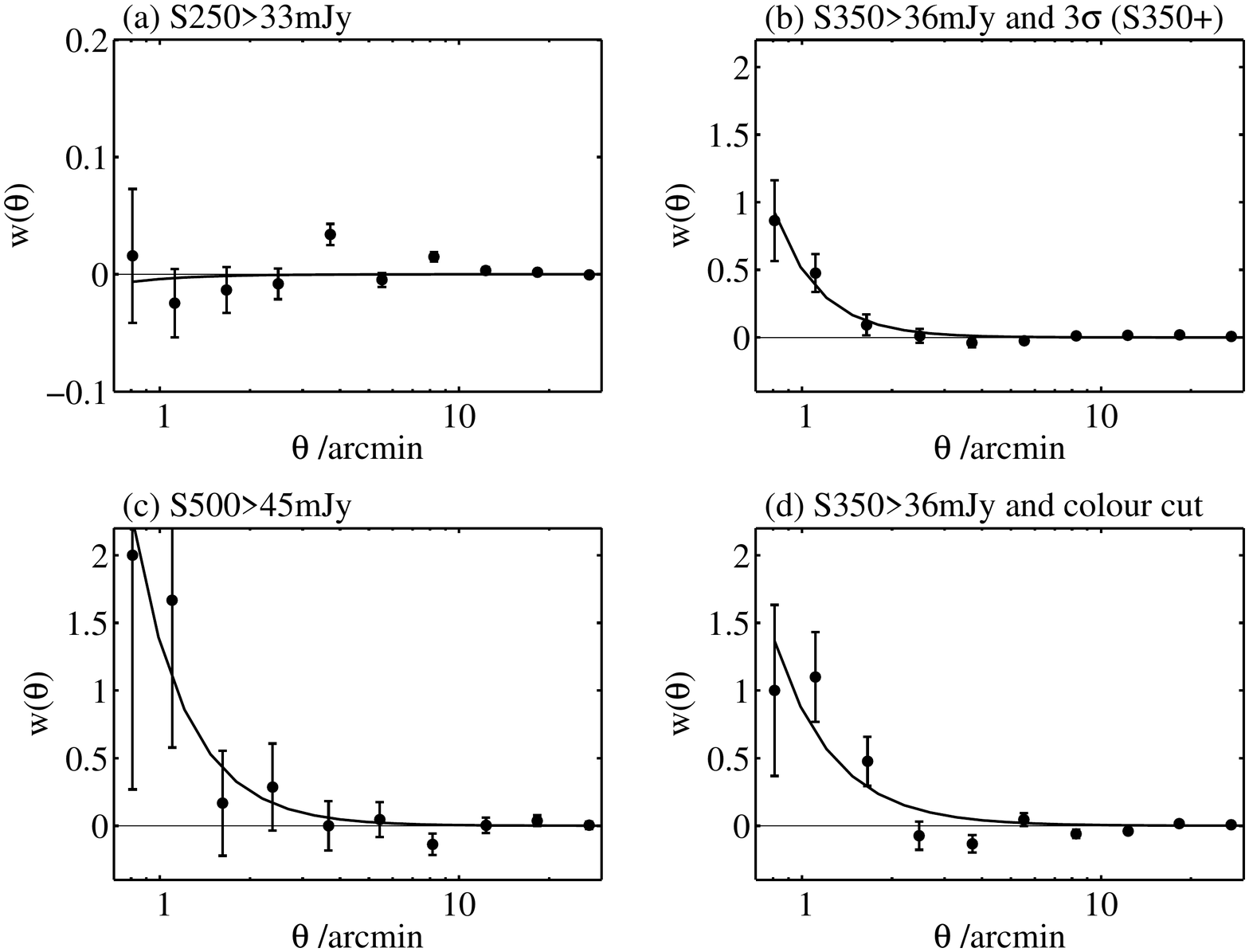}}
% figure caption is below the figure
 \caption{Plot of $w(\theta)$ for four subsamples discussed in the
   text. The panels show (a) $S_{250}>33$mJy, (b) $S_{350}>36$mJy with
   $3\sigma$ detections in the 250$\ \mu$m and 500$\ \mu$m bands
   (the 350+ sample), (c)
   $S_{500}>45$mJy, and (d) $S_{350}>36$mJy with $3\sigma$ and
   $S_{500}/S_{250}>0.75$ colour selected sample. The error bars on
   the plots are estimated from the Poisson noise in the pair
   counts. Note the expanded scale on panel (a). }
\label{fig:3}       % Give a unique label
\end{figure*}

Our measurements of $w(\theta)$ are shown in Fig.~\ref{fig:3}.  The
panels (a) and (c) show the flux limited samples in the 250 and 500$\
\mu$m, bands while panel (b) shows the $350+$ sample and panel (d) the
$S_{500}/S_{250}>0.75$ colour selected sample. The error bars on
the plots are estimated from the Poisson noise in the pair counts. 
We fitted the data using a simple power law of the form $w(\theta)
= A \theta^{-\delta} $, corrected for the integral constraint using
the Roche and Eales (\cite{integ}) technique.  The power law slopes,
$\delta$ and amplitudes at 1 arc minute, $A$ are given in
Table\ref{tab:1}.  Uncertainties on these measurements were estimated
by fitting power laws to Monte-Carlo realizations of the data, and
measuring the standard deviation of the resulting parameters.

%__________________________________________________________________
\section{Discussion }
\label{sec:4}

The $S_{250}$ sample has no detectable clustering signal. The simple
$S_{350}$ flux limited sample does show fairly significant clustering
at scales $\lesssim 2'$, but the $350+$ sample produces a higher
amplitude and more significant detection, as expected given that
adding cuts in the other two bands leads to a narrower $n(z)$ by
removing low-z galaxies.  The $S_{500}$ sample gives a noisier
$w(\theta)$, but also shows a high amplitude. The colour-selected
sample with $S_{500}/S_{250}>0.75$ shows a higher amplitude, and
higher noise compared to the $350+$ sample. The power-law fits to all
samples give steep slopes $\delta \sim 2$.

The increasing amplitude in samples selected at longer wavelengths
suggests that clustering is stronger in the higher redshift
populations, since selection at longer wavelengths tends to favour
higher redshift galaxies. According to some models, these are highly
clustered star-burst galaxies which are the ancestors of present day
ellipticals (Negrello et al. \cite{mattia}, Dav\'e et
al. \cite{dave}). The two samples with additional color cuts have
$n(z)$ from photometric redshifts derived by Amblard et
al. 2010. which can be used to convert the angular amplitudes to
spatial $r_0$.  For both samples we find a range of $r_0 \sim
7-11h^{-1}$ Mpc, depending on the slope used ($\delta=2$ or
$\delta=0.8$ respectively). While the slope we measure at scales of a
few arcmin is universally steep we cannot be sure with the current
data-set what the behaviour will be at larger scales. A larger data
set is required to fully address the behaviour of the slope and hence
reduce the uncertaintites in $r_0$.

At first sight, the non-detection of clustering in the 250$\ \mu$m
sample is somewhat surprising. It contains a high fraction of lower
redshift galaxies, with $>30$ percent at $z<1$ (Smith et
al. in prep). This low-z population is expected to cluster in a similar
way to local optical galaxies. Galaxies in the SDSS with magnitudes
$21<r^*<22$ have a correlation amplitude of $0.046$ at $1'$ (Connolly
et al. 2002). Also most of the 350$\ \mu$m sample is a subset of the
250$\ \mu$m sample, so their clustering will contribute to the
250$\ \mu$m clustering. Assuming they are uncorrelated with the low-z
galaxies, they will contribute an amplitude scaled down by the
relative density squared, leading to an expected amplitude $\sim
0.03$. These relatively small amplitudes are consistent with our
measurement of $-0.01\pm 0.07$.

Overall, these results are consistent with the general expectation
from models which include highly clustered high redshift and weakly
clustered low redshift populations of sub-mm emitting galaxies. Models
which postulate that sub-mm sources have a higher mass IMF than normal
(Baugh et al. 2005, Lacey et al. 2010) predict a higher clustering
strength at lower redshifts, which seems at first glance to be at odds
with our result. However, these models do predict far stronger
clustering above a threshold in luminosity which is redshift
dependent. It is possible that our high-z samples exceed this
threshold while the low-z samples do not. It is beyond the scope of
the current data set to be able to confirm or rule out these models.

There are few previous observations to compare to directly, and those
that are available have rather different selections and redshift
distributions.  Magliocchetti et al. (\cite{manuela}) analysed the
distribution of bright 24$\ \mu$m sources with faint optical
counterparts, split into high ($\langle z \rangle \sim 2$) and low
redshift ($\langle z \rangle \sim 0.8$) sub-samples. They found a low
amplitude ($A\sim 0.14 \pm 0.05 $) at low redshift and a higher
amplitude ($A\sim 0.26\pm0.1$) at high redshift. This trend of
$w(\theta)$ increasing towards higher z is similar to our
observations.

The power spectrum analysis of the BLAST data by Viero et al.
(\cite{viero}) detects clustering on scales $5' < \theta < 20'$.  
Once interpreted within the Halo Model formalism, their measurement
points to an increase in the spatial clustering of the background
sub-mm source population with increasing wavelength and therefore
increasing redshift. Again this is consistent with our findings.

LABOCA observations of the extended Chandra Deep Field South have
produced a catalogue of 126 sources selected at 850$\ \mu$m
(Wei\ss\ et al \cite{weiss}). A weak detection of clustering is found
on scales $\theta<2'$. Fixing the slope to be $0.8$, their power-law
fit has an amplitude of $0.18\pm 0.1$ at $1'$. This is similar to the
amplitude that we find for the 350$\ \mu$m selected sample.

Though all three of these measurements are similar to ours, it is not
simple to make a direct comparison because either or both of the
flux-limit or pass-bands are different, and so the redshift
distributions are not the same.

Given the statistical limitations on the current small field, we leave
detailed comparisons to models to a later analysis with more
data. However we can say that our measurements appear to be consistent
with the prevailing models, where sub-mm sources consist of a low
redshift population of normal star-forming disk galaxies that have a
spatial correlation length, $r_0 \sim 4$ Mpc, and a high redshift
population of highly clustered star-bursting galaxies $r_0 \sim 10$Mpc
(Negrello et al. \cite{mattia}, Narayanan et al. \cite{narayanan},
Dav\'e et al. \cite{dave}). The rather steep slope of our measurements
for the higher redshift samples is also consistent with the high
redshift population forming in compact protoclusters.

The current field is only 1/30 of the area that H-ATLAS will cover,
and it has the brightest cirrus background of the planned fields. The
final H-ATLAS dataset will have much larger fields with lower cirrus
emission, and so will provide a benchmark measurement of source
clustering in sub-mm populations.

% 1'= 0.5085Mpc at z=2

%__________________________________________________________________
%\begin{acknowledgements}
%\end{acknowledgements}

\end{document}